\documentclass[twocolumn,showpacs,preprintnumbers,amssymb,amsmath,superscriptaddress,letterpaper]{revtex4}[10pt]

\usepackage{graphicx}
\usepackage{dcolumn}
\usepackage{bm}
\usepackage{pstricks}
\usepackage{color}
\usepackage{epsfig}

\newcommand{\be}{\begin{equation}}
\newcommand{\ee}{\end{equation}}
\newcommand{\bea}{\begin{eqnarray}}
\newcommand{\eea}{\end{eqnarray}}

\def\ltap{\ \raise.3ex\hbox{$<$\kern-.75em\lower1ex\hbox{$\sim$}}\ }
\def\gtap{\ \raise.3ex\hbox{$>$\kern-.75em\lower1ex\hbox{$\sim$}}\ }

\begin{document}

\title{Charge Asymmetric Cosmic Ray  Signals From Dark Matter Decay}

\author{Spencer Chang}
\affiliation{Physics Department, University of California Davis, Davis, CA 95616}\affiliation{Department of Physics, University of Oregon, Eugene, OR 97403}

\author{Lisa Goodenough}
\affiliation{Center for Cosmology and Particle Physics, Department of Physics, New York University, 
New York, NY 10003}

\date{\today}

\begin{abstract}

The PAMELA and Fermi measurements of the cosmic-ray electron and positron spectra have generated much interest over the past two years, because they are consistent with a significant component of the electron and positron fluxes between 20 GeV and 1 TeV being produced through dark matter annihilation or decay.  However, since the measurements are also consistent with astrophysical interpretations, the message is unclear.  In this paper, we point out that dark matter can have a more distinct signal in cosmic rays, that of a charge asymmetry.  Such charge asymmetry can result if the dark matter's abundance is due to a relic asymmetry, allowing its decay to generate an asymmetry in positrons and electrons.  This is analogous to the baryon asymmetry, where decaying neutrons  produce electrons and not positrons. We explore benchmark scenarios where the dark matter decays into a leptophilic charged Higgs boson or electroweak gauge bosons.  These models have observable signals in gamma rays and neutrinos, which can be tested by Fermi and IceCube.  The most powerful test will be at AMS-02, given its ability to distinguish electron and positron charge above 100 GeV.   Specifically, an asymmetry favoring positrons typically predicts a larger positron ratio and a harder (softer) high energy spectrum for positrons (electrons) than charge symmetric sources.  We end with a brief discussion on how such scenarios differ from the leading astrophysical explanations.

\end{abstract}
\maketitle

\section{Introduction}
\label{sec:intro}
The consistent observations of dark matter (DM) on a wide range of scales have made DM a cornerstone of the standard cosmological model.  Yet to this day, its fundamental origin remains a mystery.  There is hope that this will change soon.  For WIMP (weakly interacting massive particle) dark matter, signals are expected at colliders and also direct and indirect detection experiments \cite{Jungman:1995df}.  However, confirmation of such signals is not straightforward.  This is best demonstrated by indirect detection experiments, where observations of dark matter-like signals in antiparticles, neutrinos, and gamma rays may be instead due to astrophysical sources.  In fact, even though PAMELA and Fermi have observed excesses in positrons and electrons that are consistent with a DM interpretation, the conclusion is unclear since there are competing astrophysical explanations.  For this reason, it is worth considering a broader range of dark matter signals to determine if dark matter can have distinctive characteristics when compared to astrophysical hypotheses.  In this paper we discuss one specific example, which we believe merits further investigation.  

As has been discussed in numerous papers, the positrons and electrons observed by PAMELA \cite{Adriani:2010ib}, Fermi \cite{Abdo:2009zk}, and HESS \cite{Aharonian:2008aa,Aharonian:2009ah} are consistent with dark matter scenarios requiring unconventional dynamics.  The Fermi and HESS electron spectra point to heavy dark matter with a mass $m_{\chi}\gtrsim$ 1 TeV.  This, coupled with the rise in the positron fraction measured by PAMELA, has generated the most interest in two broad classes of dark matter.  The first class has annihilating dark matter with an annihilation cross section $O(100-1000)$ times larger than the relic abundance expectation \cite{Cirelli:2008pk, Cholis:2008hb}.  This has motivated new dynamics to achieve such high cross sections, for e.g. Sommerfeld enhancement through a light mediator \cite{ArkaniHamed:2008qn}.    The second class, decaying dark matter, requires dark matter decay lifetimes longer than the age of the universe, of order $10^{26}$ seconds (for some of the earliest references see \cite{Chen:2008yi, Chen:2008dh, Yin:2008bs, Ishiwata:2008cv}).   Despite the difference in the underlying dynamics, the two scenarios can give the same electron and positron signals, provided the annihilation products of the first class are the same as the decay products of the second class.  Neutrino and gamma-ray fluxes, which are sensitive to the difference between DM density (decay) and DM density squared (annihilation), have potentially more discriminating power.  

Many astrophysical explanations of the electron excesses have also been proposed.  The most prominent are pulsar explanations \cite{Hooper:2008kg, Profumo:2008ms}, but other mechanisms exist, including secondary positrons and electrons produced in the acceleration mechanism of supernovae remnants \cite{Blasi:2009hv, Ahlers:2009ae}, and local inhomogeneities in supernovae remnants \cite{Piran:2009tx}.  Given the uncertainties in the modeling, these sources can provide adequate fits to the positron and electron excesses.  There is hope to distinguish a dark matter explanation from these astrophysical explanations by looking for additional signals in neutrinos and gamma rays, given the additional directional information available in these signals.  However, even using gamma rays to distinguish between dark matter decays and annihilations has been brought into question \cite{Boehm:2010qt}, so it is not clear whether these would provide concrete conclusions.  Also, in some dark matter models, the gamma-ray and neutrino signals may not give observable rates, making it difficult to give a correlated confirmation.      

In this paper, we consider a relatively unexplored possibility in the dark matter signal, that of charge asymmetry \footnote{See \cite{Feldstein:2010xe}, for a recent paper discussing asymmetric signals in neutrinos versus antineutrinos.}.  Such charge asymmetry gives a potential discriminant against the astrophysical alternatives.   For instance, pulsars are expected to produce equal contributions of electrons and positrons with no appreciable charge asymmetry, while a charge asymmetry can occur naturally in the context of dark matter decays.  In particular, dark matter that acquires its abundance due to a relic asymmetry can have decays that produce charge asymmetries.  An example of this occurs in the baryon sector, where neutron beta decay produces electrons but no positrons.  Such a signal has not yet been considered for dark matter.  Annihilations occur in a charge neutral state and thus cannot produce charge asymmetries without violation of charge conjugation (C) symmetry.  So far, most decaying models have not considered the required dark matter asymmetry (or a significant C asymmetry in the decay).  In this paper, we present six benchmark models of charge asymmetric dark matter.  We discuss the distinguishing signals of charge asymmetry and how they might be probed at future experiments.    

\section{Analysis}
\label{sec:analysis}
As pointed out before (see e.g. \cite{Barr:1990ca, Nardi:2008ix}), dark matter with a relic asymmetry can naturally have TeV scale masses, due to an exponential suppression of the asymmetry.  A well known example is a technibaryon that has electroweak constituents, so that sphalerons transfer an asymmetry between technibaryons and baryons.  In this case, the relic asymmetry of technibaryons (TB) is related to baryons (B) by \cite{Barr:1990ca}
\bea
\nonumber
TB \sim B \; e^{-m_{TB}/T_{sphaleron}},
\eea
where $T_{sphlaeron}$ is the freeze out temperature of sphalerons, which is expected to be $O(100)$ GeV, allowing masses $m_{TB} \sim $ TeV with the correct dark matter relic abundance.  Thus, sphalerons, and more generally any other asymmetry transferring process which decouples at the weak scale, can naturally accommodate TeV scale asymmetric dark matter.  Furthermore, in many cases, the surviving dark matter has baryon minus lepton number equal to one, leading to a large class of models in which dark matter has the quantum numbers of an antilepton, which naturally leads to more positrons than electrons.   Thus, asymmetric dark matter seems to have the generic ingredients for a charge asymmetry that can explain the electron/positron data.  Asymmetric dark matter has a long history (see \cite{Nussinov:1985xr, Barr:1990ca, Harvey:1990qw, Kaplan:1991ah}) and has had a recent surge in interest due to the models described in \cite{Kaplan:2009ag}.  For discussion of TeV mass asymmetric dark matter, see \cite{Buckley:2010ui, Shelton:2010ta, Falkowski:2011xh}, and for work on the indirect detection signals of asymmetric dark matter, see \cite{Cohen:2009fz, Cai:2009ia, Khlopov:2009hi, Feldstein:2010xe}.

We now turn to a set of benchmark models to illustrate the potential signals, leaving detailed explorations of models and other extensions to future work.  We consider the case of a dark matter fermion $\chi$, which is not equal to its own antiparticle $\bar{\chi}$.  Our assumption is that this sector has an asymmetry such that, after DM -- anti-DM annihilations in the early universe, essentially only $\chi$'s exist today.   We assume that $\chi$ has antilepton number, giving rise to the well-motivated decays
\bea
\nonumber
\chi &\to& X^-\: (\mu^+,\tau^+), \\[-.2cm]
\\[-.2cm]
\nonumber
\chi &\to& X^0\: \bar{\nu}.
\eea
We only consider decays to $\mu$ and $\tau$, because decays to electrons give a peaked structure in the electron spectrum not seen in the Fermi and HESS data.  The signals for our scenarios depend on the choice of the particles $X$.  In the standard model, the options are $X^- = W^-$ and $X^0 = Z^0, h^0$.  The benchmarks in this case are 
\bea
&i)& W^- \ell^+  \; \text{decays only,} \label{eqn:decays0}\\
&ii)& W^- \ell^+, Z^0 \bar{\nu}, h^0 \bar{\nu} \; \text{with branching ratios 2:1:1.} \label{eqn:decays1}
\eea
Case $ii)$ reflects decays similar to a right-handed neutrino.  We find that the antiproton constraint from PAMELA \cite{Adriani:2010rc} is in tension with these two cases (see the discussion in Section~\ref{sec:constraints}), thus we consider a variant that is, by construction, safer \footnote{We do not consider final state radiation of electroweak bosons as pointed out by  \cite{Ciafaloni:2010ti, Bell:2010ei}.  Such higher order effects will lead to a high energy component of antiprotons, but that should still be consistent with PAMELA.}.  In particular, we look at the scenario
\bea
&iii)& \chi \to H^- \ell^+ \;\; \text{where} \;\; H^- \to \tau^- \bar{\nu},\label{eqn:decays2}
\eea
i.e.~where $H^-$ is a leptophilic charged Higgs dominantly decaying to taus.  Such a Higgs sector has already been discussed in the context of explaining PAMELA's positron fraction \cite{Goh:2009wg}, but without considering a charge asymmetry.  Note that this benchmark is an optimistic case which will lead to a larger charge asymmetry.  We discuss some of the model building issues for these benchmarks in the appendix, leaving for future work their full realization in a complete theory.

For our benchmark models, we set the mass of the charged Higgs $H^-$ to be 150 GeV and of the dark matter $\chi$ to be 3.5 (7.0) TeV for muon (tau) decays.  The $\chi$ mass is chosen to give an electron spectrum that agrees with the high energy softening seen at Fermi \cite{Abdo:2009zk} and HESS \cite{Aharonian:2008aa,Aharonian:2009ah}.  The signals are only weakly dependent on the charged Higgs mass, as long as it is not comparable to the mass of $\chi$.

\begin{figure}[t]
\begin{center}
\vskip -1.65in
\epsfig{figure=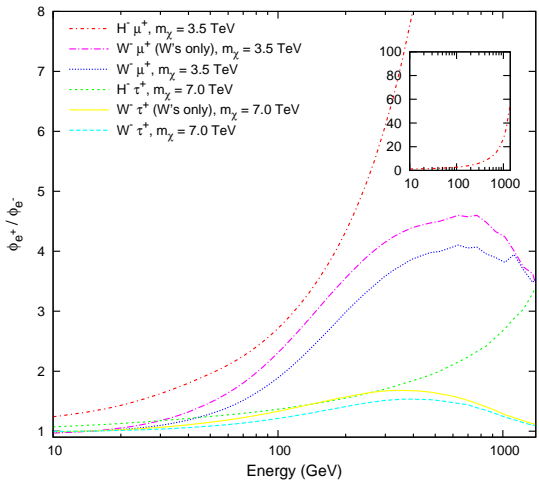,width=0.42\textwidth}
\end{center}
\caption{The local dark matter contribution to the ratio of fluxes $\Phi_{e^+}/\Phi_{e^-}$ for all of the benchmark models.  The inset figure shows a zoomed out plot for the $H^- \mu^+$ model.}
\label{fig:electronpositronratio}
\end{figure}

In Figure \ref{fig:electronpositronratio}, we show the local dark matter contribution to the ratio of fluxes $\Phi_{e^+}/\Phi_{e^-}$ after cosmic ray propagation \endnote{We use GALPROP \cite{galprop} to calculate all cosmic ray and gamma ray fluxes.}, demonstrating that the charge asymmetry can be significant.  Due to the hard antileptons produced in the decays of Eqns.~\ref{eqn:decays0}-\ref{eqn:decays2}, the ratio increases as a function of energy.  In general, the muon models have larger charge asymmetries than the tau models, while the charged Higgs decays have a larger asymmetry than the electroweak gauge boson decays.  Accordingly, the muon models will have more distinctive consequences due to the enhanced charge asymmetry.  In a recent paper \cite{Frandsen:2010mr}, new sources of electrons and positrons with constant charge asymmetry were discussed and constrained by ``cosmic sum rules."  However, only extreme asymmetries were disfavored, which our models avoid in the relevant energies below 100 GeV. 

\begin{figure}[t]
\begin{center}
\vskip -0.20in
\epsfig{figure=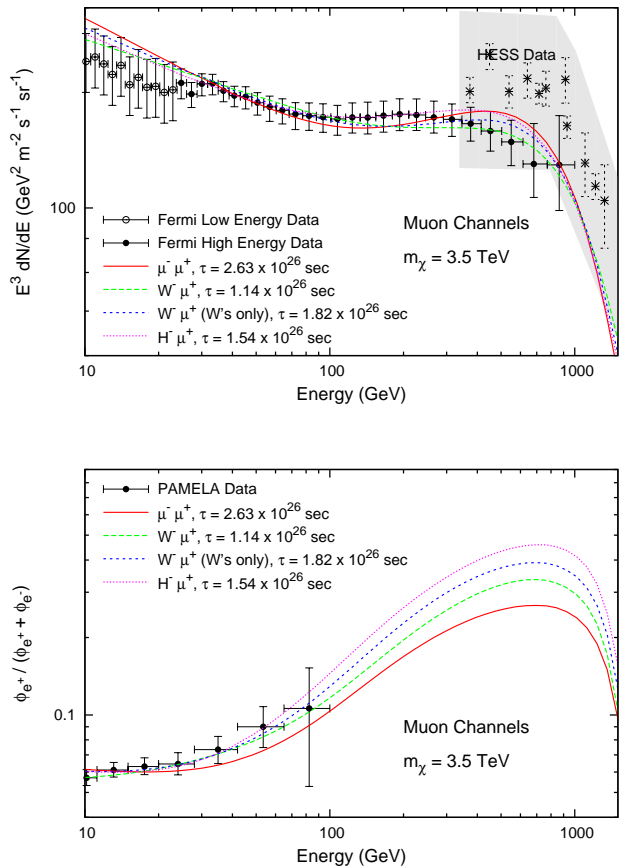,width=0.48\textwidth}
\end{center}
\caption{The fits for electrons+positrons (\emph{top}) and the positron fraction (\emph{bottom}) for the muon models and a charge symmetric case $\chi \to \mu^+ \mu^-$.}
\label{fig:muonfits}
\end{figure}

\begin{figure}[t]
\begin{center}
\epsfig{figure=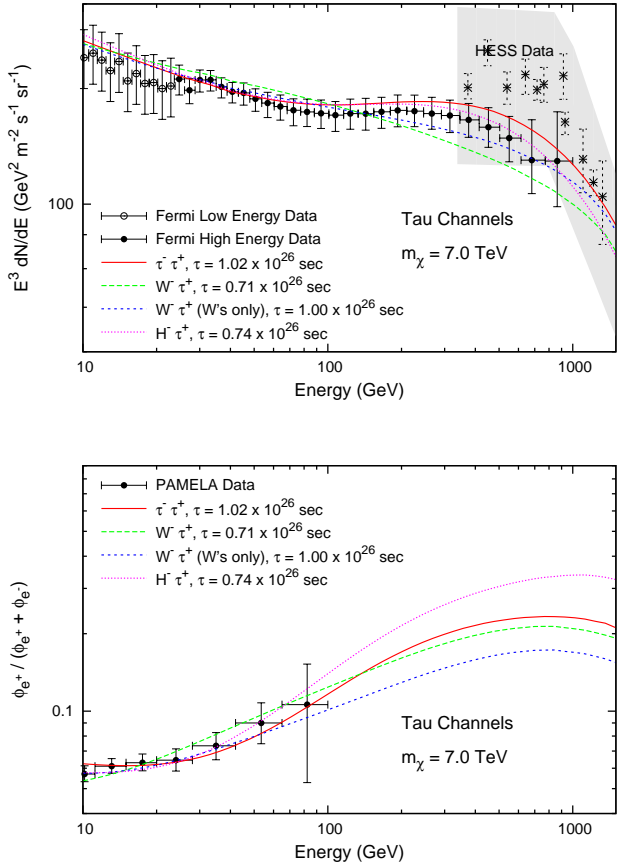,width=0.48\textwidth}
\end{center}
\caption{The fits for electrons+positrons (\emph{top}) and the positron fraction (\emph{bottom}) for the tau models and a charge symmetric case $\chi \to \tau^+ \tau^-$.}
\label{fig:taufits}
\end{figure}

We fit to the Fermi and HESS electron+positron data and PAMELA positron fraction data for all six models and show the resulting spectra in Figure~\ref{fig:muonfits} (muon models) and Figure~\ref{fig:taufits} (tau models).  For comparison, we include the charge symmetric models, $\chi \to (\mu^+ \mu^-, \tau^+ \tau^-)$.   Typically in charge symmetric decaying dark matter scenarios, PAMELA requires a shorter lifetime than Fermi.  As can be seen in the figures, the charge asymmetry resolves this normalization inconsistency between PAMELA and Fermi.    If the asymmetry favors positrons over electrons, we see that there is a consistent dark matter signal which gives good fits to both data sets.  The figures also demonstrate that an asymmetry can lead to a larger positron fraction at high energies, after fitting to the Fermi and HESS spectra.  A smoking-gun signal for a new, charge asymmetric source would exist if the positron fraction were to exceed 1/2.  Unfortunately, for the benchmark models we consider this value is not quite reached.  However, it could occur in other models.  As a final comment, although the visually striking differences in the positron fraction occur at high energies ($\gtrsim$ 100 GeV), the differences at low energy are large enough that improved measurements in this energy range could help distinguish these scenarios.

\begin{figure}
\begin{center}
\includegraphics[width=.42\textwidth]{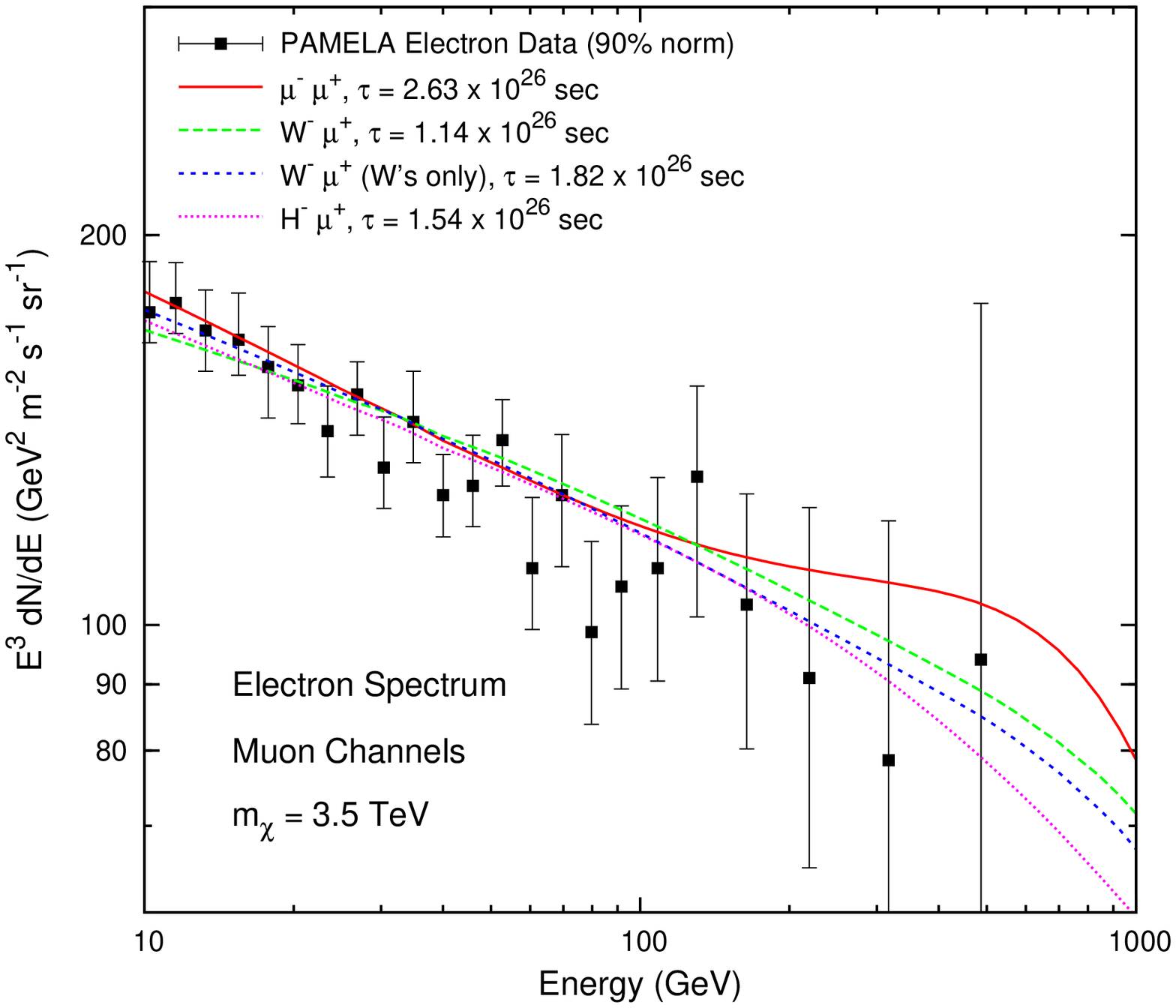}\\
\vskip 0.05in
\includegraphics[width=.42\textwidth]{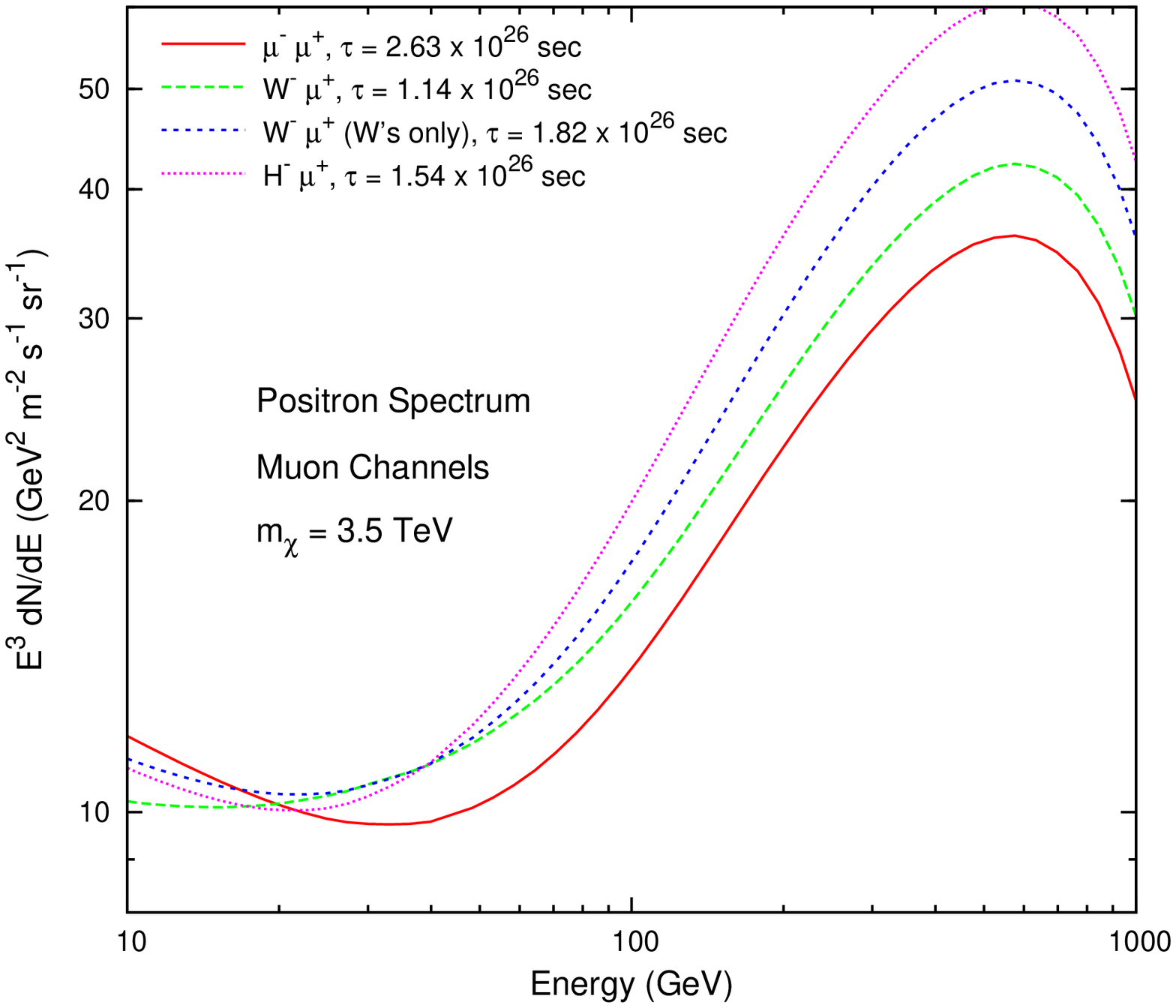}
\end{center}
\caption{The electron (\emph{top}) and positron (\emph{bottom}) spectra for the asymmetric muon models and a charge symmetric case $\chi \to \mu^+ \mu^-$. The electron spectrum data are the recently released data from PAMELA \cite{Adriani:2011xv} normalized to be consistent with the Fermi electronic data.  The behavior of the electron spectrum for energies 100-400 GeV for the charge symmetric model (\emph{solid red line}) is $dN/dE \propto E^{-3.10}$.  For the charge asymmetric models introduced here, the power law behavior is noticeably softer.}
\label{fig:muonepem}
\end{figure}

\begin{figure}
\begin{center}
\includegraphics[width=.42\textwidth]{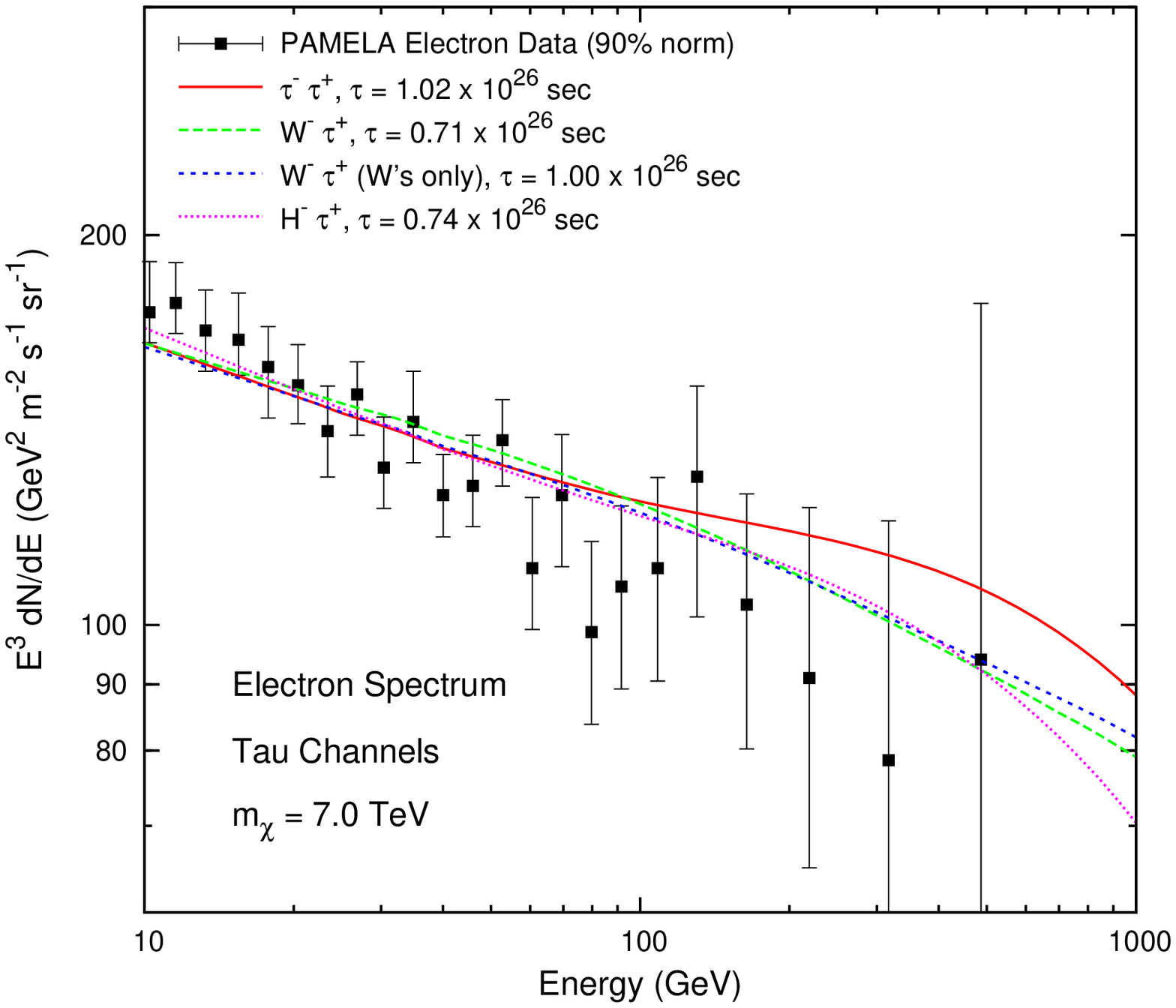}\\
\vskip 0.05in
\includegraphics[width=.42\textwidth]{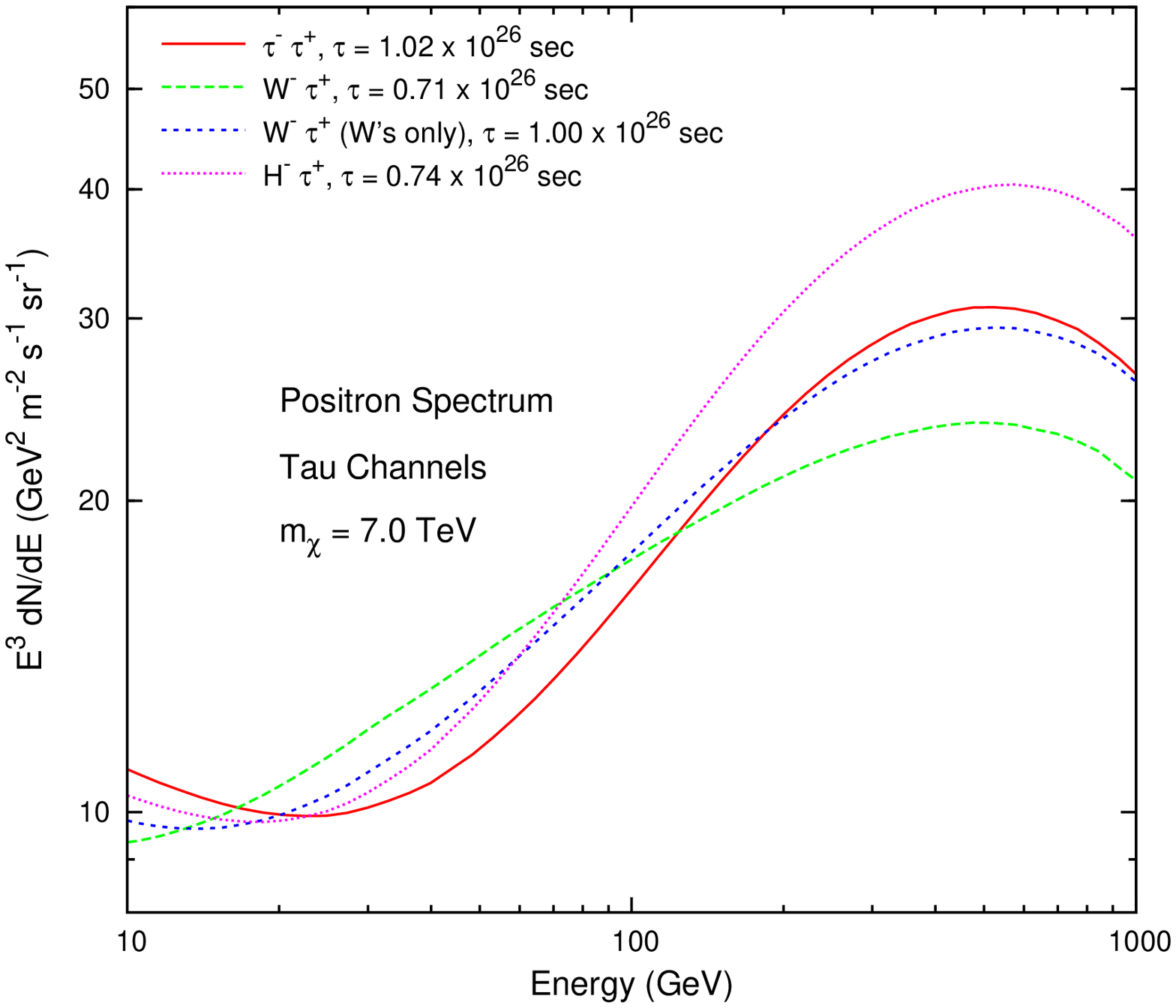}
\end{center}
\caption{The electron (\emph{top}) and positron (\emph{bottom}) spectra for the asymmetric tau models and a charge symmetric case $\chi \to \tau^+ \tau^-$.  The electron spectrum data are the recently released data from PAMELA \cite{Adriani:2011xv} normalized to be consistent with the Fermi electronic data.  The behavior of the electron spectrum for energies 100-400 GeV for the charge symmetric model (\emph{solid red line}) is $dN/dE \propto E^{-3.09}$.  For the charge asymmetric models introduced here, the behavior is $dN/dE \propto E^{-3.16}$.}
\label{fig:tauepem}
\end{figure}

Of course, there is more information to be gained by looking at the electron and positron spectra separately.  As shown in Figure~\ref{fig:muonepem} (muons) and Figure~\ref{fig:tauepem} (taus), the electron and positron spectra can give additional evidence that the source of electrons and positrons is charge asymmetric.  Included is PAMELA's electron spectra \cite{Adriani:2011xv}, based on three-and-a-half years of data, which has been normalized to be consistent with the Fermi electronic data.   A charge asymmetry favoring positrons gives rise to an electron spectrum that is considerably softer at high energies than that of a charge symmetric scenario.  For the muon channels we consider, the behavior of the electron spectrum (see the top of Figure~\ref{fig:muonepem}) between 100 and 400 GeV can be described by a power law, $dN/dE \propto E^{-\alpha}$, with $\alpha \approx 3.10$ for the charge symmetric case and $\alpha \gtrapprox 3.19$ for the charge asymmetric cases.  The differences in the electron spectra for the tau channels are less pronounced (see the top of Figure~\ref{fig:tauepem}).  Between 100 and 400 GeV, the charge symmetric case has power law behavior with $\alpha \approx 3.09$, while the charge asymmetric cases have $\alpha \gtrapprox 3.16$.

The differences in the positron spectra at high energy are not as straightforward.  For the muon channels (see bottom of Figure~\ref{fig:muonepem}), the power law behavior between 100 and 300 GeV is described by $\alpha \lessapprox 2.21$ for the charged Higgs decay channel and the $W^- \mu^+$ only channel.  However, both the \emph{charge asymmetric} $W^- \mu^+, Z^0 \bar{\nu}, h^0 \bar{\nu}$ (with branching ratios 2:1:1) channel and the \emph{charge symmetric} $\mu^+ \mu^-$ decay channel have $\alpha \approx 2.29$ between 100 and 300 GeV.  Although these two channels have very similar power law behaviors above 100 GeV, their low energy behaviors are quite different; between 10 GeV and 20 GeV the $W^- \mu^+, Z^0 \bar{\nu}, h^0 \bar{\nu}$ channel has $\alpha \approx 3.03$ and the $\mu^+ \mu^-$ channel has $\alpha \approx 3.23$.  

\newpage

The situation is much the same for the tau channels at high energies (see Figure~\ref{fig:tauepem}), though here it is the charged Higgs and the symmetric decay channels that have similar power law behaviors for positrons above 100 GeV; the $W^- \mu^+$ only and $W^- \mu^+, Z^0 \bar{\nu}, h^0 \bar{\nu}$ channels are noticeably softer.  The $H^+ \tau^-$ channel and charge symmetric $\tau^+ \tau^-$ channel are not only similar at high energies, but also at low energies (10-20 GeV) with $\alpha \approx 3.12$ ($H^+ \tau^-$) and $\alpha \approx 3.17$ ($\tau^+ \tau^-$).

Distinguishing these behaviors in the electron and positron spectra naturally relies on charge discrimination.  As can be seen in the figures, PAMELA's electron data  \cite{Adriani:2011xv}, which go up to 625 GeV, have only a slight preference for charge asymmetry, but do not have sufficient statistics above 100 GeV to convincingly differentiate the scenarios.  AMS-02, which was just recently launched, is projected \cite{Kounine:2010js} to be capable of precisely probing the positron fraction as well as the electron and positron spectra in an energy range $\sim 10 - 800$ GeV, enabling the features described to be tested.          

\section{Constraints and Additional Signals}
\label{sec:constraints}
\begin{figure}
\begin{center}
\includegraphics[width=.48\textwidth]{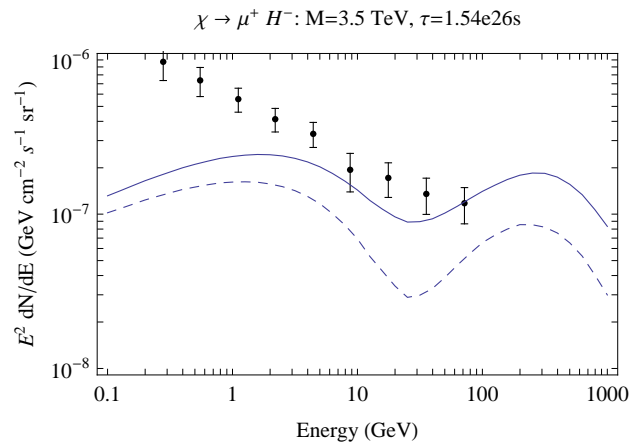}
\end{center}
\caption{The isotropic gamma-ray flux as measured by the Fermi-LAT shown along with the $H^- \mu^+$ model's Galactic anticenter + extragalactic contribution (\emph{solid}), and extragalactic contribution alone (\emph{dashed}).}
\label{fig:isotropic}
\end{figure}

Charge asymmetric dark matter decay models have additional signals and constraints.  In particular, gamma-ray data from Fermi can place strong constraints on DM decays \cite{Chen:2009uq,Cirelli:2009dv,Papucci:2009gd}.  For one, the extragalactic isotropic component as measured by Fermi \cite{Abdo:2010nz} is a potential constraint.  In Figure~\ref{fig:isotropic} we show the dark matter contribution to the isotropic gamma-ray flux for the $H^- \mu^+$ model, along with the Fermi observation; we see that they are consistent.  Unlike other decay modes, adding the smallest galactic component, i.e. that from the Galactic anticenter, to the extragalactic contribution is still allowed \cite{Cirelli:2009dv}.   For the other benchmarks, the constraints are tighter.  For the $W^- \mu^+$ and $H^- \tau^+$ models, the gamma ray spectra is consistent if the minimal galactic component is not added; with the galactic contribution added, the $W^- \mu^+$ models are ruled out and the $H^- \tau^+$ is  about $1\sigma$ high on the last data point.  Finally, the $W^- \tau^+$ models are ruled out considering only the extragalactic component alone.  These issues suggest that as Fermi accumulates data above 100 GeV for their isotropic measurement, they should be sensitive to (or further exclude) these scenarios.       

Gamma-ray flux limits from galaxy clusters are also a potential constraint \cite{Ackermann:2010qj}.  As shown in \cite{Dugger:2010ys}, clusters like Fornax can place significant bounds on decaying dark matter.  In this case, the charge asymmetry actually helps relax these constraints.  The strongest flux limits for the clusters are in the 1-10 GeV range \cite{Ackermann:2010qj}.  Gamma rays in this energy range arise due to inverse Compton scattering between the cosmic microwave background and the $\gtrsim 500$ GeV electrons and positrons injected by dark matter decay.  Due to the charge asymmetry, for the same dark matter lifetime, there are roughly half as many high energy $e^- + e^+$ for the charge asymmetric model as compared to the symmetric model.  For example, in the decay $\chi \to H^- \mu^+$ there is only one energetic antimuon produced, while in $\chi \to \mu^+ \mu^-$ both an energetic muon and antimuon are produced.  Thus, in the relevant range, the limits are weakened by about 50\%.  Reproducing the analyses in \cite{Ackermann:2010qj, Dugger:2010ys}, we find that the models are not ruled out by more than a factor of 1.57 by Fornax (the most constraining cluster), which is still within the uncertainty of total dark matter mass determinations of Fornax \cite{Dugger:2010ys}.  It is also worth pointing out that these benchmarks predict a larger rate for high energy gamma rays (due to photons produced in tau decays), which are not constrained at the moment.  Thus, future Fermi gamma-ray observations of clusters could see some indications of gamma rays in the 1-10 GeV range and potentially at higher energy.

Neutrinos are another possible source of constraints and future signals for asymmetric dark matter models.  In particular, $\chi \to X^0 \bar{\nu}\,$ decays give a distinctive, monochromatic antineutrino signal.  Similar to the charge asymmetry, there is also an asymmetry between neutrinos and antineutrinos in these models, which through charged current interactions produces a charge asymmetry in their products.   Such distinctions, for example in muon and antimuon fluxes, would be hard to detect at neutrino telescopes \footnote{However, low energy neutrino and antineutrino discrimination has been argued to be detectable at detectors like MINOS for asymmetric dark matter models \cite{Feldstein:2010xe}.} due to the small deflection in the Earth's magnetic field.  For all of these signals, we find from recent analyses \cite{Mandal:2009yk,Covi:2009xn,Falkowski:2009yz} that the limits from Super-K and AMANDA are not sensitive to the lifetime of the benchmark model.  However, future analyses at ANTARES \cite{:2010ec} and IceCube \cite{Achterberg:2006md} should be at least sensitive to this model at 90\% CL. 

\begin{figure}[t]
\begin{center}
\epsfig{figure=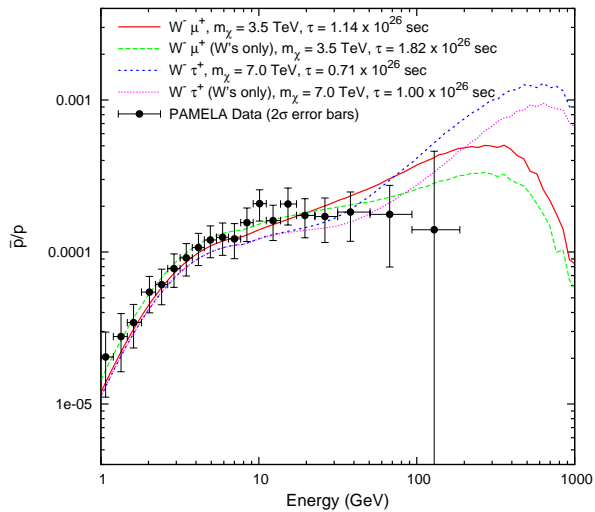,width=0.45\textwidth}
\end{center}
\caption{The antiproton-to-proton flux ratio $\Phi_{\bar{p}}/\Phi_{p}$ for the $W^- \ell^+$ and $W^- \ell^+, Z^0 \bar{\nu}, h^0 \bar{\nu}$ (2:1:1) decay channels.  The PAMELA data for the ratio are shown with 2$\sigma$ error bars.}
\label{fig:antiprotons}
\end{figure}

The PAMELA measurements of the antiproton flux and antiproton-to-proton flux ratio up to 180 GeV \cite{Adriani:2010rc} place constraints on any dark matter model with antiprotons as decay products.   Because we consider charge asymmetric decays in which only one $W$ or $Z$ boson is created in each decay, rather than two as is the case in charge symmetric DM decays, our models have half the antiproton production of conventional dark matter decay models with the same DM lifetime.  Moreover, the asymmetry in the positron and electron fluxes (with the relative enhancement in the positrons) allows for the Fermi and PAMELA electron/positron data to be fit with a smaller dark matter decay rate.  These differences are enough to prevent our models from being ruled out by the current antiproton constraints, unlike conventional DM decay modes.  For the DM lifetimes needed to fit the Fermi and PAMELA electron/positron data as shown in Figures~\ref{fig:muonfits} and~\ref{fig:taufits}, the $W^- \mu^+$, $W^- \tau^+$, and $W^- \mu^+, Z^0 \bar{\nu}, h^0 \bar{\nu}$ (2:1:1) modes are consistent at the $2 \sigma$ level, while the $W^- \tau^+, Z^0 \bar{\nu}, h^0 \bar{\nu}$ (2:1:1) mode is mildly inconsistent with the two highest energy data points at the $2 \sigma$ level.  See Figure~\ref{fig:antiprotons}.

As discussed in \cite{Goh:2009wg}, leptophilic Higgs sectors require Higgs masses lower than 200 GeV to avoid antiprotons from $W, Z$ decays, hence they can be pair produced at the LHC with cross sections larger than 10 fb and searched for in $H^0 H^0 \to 4 \tau$, $H^0 H^+ \to 3\tau \nu$, and $H^+ H^- \to 2\tau \nu$ events.  They can also be singly produced via gluon fusion and vector boson fusion.  However, they have typically smaller rates than the Standard Model Higgs due to their reduced contributions to the top and $W, Z$ masses.   For more discussion on the collider probes of leptophilic Higgs sectors, see \cite{Akeroyd:1994ga, Aoki:2009ha, Su:2009fz, Logan:2009uf, Davidson:2010sf}. 

\section{Conclusions}
\label{sec:conclusions}

While current experimental results are inconclusive about the presence of a high energy charge asymmetric source in the positron/electron data, future measurements with improved charge discrimination  will further test this possibility.  In particular, PAMELA continues to collect data and will measure the positron and antiproton fractions with improved statistics and at energies above those of their current data points.  AMS-02 plans to measure the positron and antiproton spectra up to several hundred GeV.  As argued, better measurement of the positron and electron spectra (at both high and low energies) will be crucial in determining if there is a significant charge asymmetry in the local electron and positron spectra.  If a significant asymmetry exists, it will be in substantial tension with the pulsar interpretation \cite{Hooper:2008kg, Profumo:2008ms}, since pulsars give charge symmetric injection.  In addition, for nearby pulsars, the pulsar explanation is already being tested by Fermi limits on electron/positron anisotropies \cite{Ackermann:2010ip}.  Furthermore, if the positron fraction continues to rise above 100 GeV, the local supernovae inhomogeneity scenario will be disfavored, since it predicts the fraction to drop  above 100 GeV \cite{Piran:2009tx}.  This would leave the possible explanation of secondaries produced in the acceleration shock \cite{Blasi:2009hv, Ahlers:2009ae} (it should be noted that a recent analysis \cite{Kachelriess:2011qv} finds that the secondary production is smaller than needed to explain the positron fraction).  This scenario predicts two additional, distinct phenomena.  It predicts that the ratios of secondaries-to-primaries, such as antiprotons-to-protons and Boron-to-Carbon, rise at higher energies \cite{Blasi:2009bd, Mertsch:2009ph, Fujita:2009wk}, while the dark matter scenario can only potentially give a rise in the antiproton fraction.  The acceleration shock scenario also predicts an injection of positrons and electrons of the same shape, which might be discernible given enough statistics.  Furthermore, although this does not occur in our benchmark models, the dark matter charge asymmetry can in principle be large enough that the positron fraction will asymptote above 1/2 at high energies, which would not only rule out most astrophysical explanations, but dark matter annihilation scenarios as well.  

It is important to remember that the physics of the dark sector can yield unexpected surprises.  In this paper, we demonstrated that in a scenario where the abundance of dark matter is due to a relic asymmetry, dark matter decays can give distinguishing new signals at indirect detection experiments, manifested by a charge asymmetry.  We provided six particularly simple examples in our benchmark models and note that there is much room for further exploration of the possibilities.  As we discussed, charge asymmetric decay models will be probed in the next few years at indirect detection experiments.  In addition, depending on the particle content of the model, they can also be tested at collider experiments like the LHC.  Thus, there is great potential in combining experiments in order to unravel the mysteries of such a dark sector.   

\vskip 0.2 in
\noindent {\bf Acknowledgments}
\vskip 0.05in
\noindent 
SC would like to thank Tesla Jeltema for discussions and Dan Phalen for comments on the manuscript and help on the model building.  SC is supported in part by the US Department of Energy under contract No. DE-FG02-91ER40674.  LG was supported by the Henry M. MacCracken Fellowship at New York University.  Additionally, LG acknowledges the Institute for Advanced Study in Princeton for their hospitality and support during the later stages of this work.

\section{Appendix:  Model Building Discussion}
\label{sec:modelbuilding}
Our starting point for the model building is a theory described by the following Lagrangian
\bea
{\cal L} = {\cal L}_{adm} + {\cal L}_{decay}\, \text{,}
\eea
which separates into two pieces.  The first piece, ${\cal L}_{adm}$, contains all of the dynamics required to transfer an asymmetry to dark matter and the annihilation which depletes the symmetric component.  The second piece, ${\cal L}_{decay}$, describes the interaction mediating the dark matter decay, which is what is relevant for the indirection detection signals.  In the following, we focus on ${\cal L}_{decay}$.  Given that it involves extremely weak couplings (for a dark matter lifetime of $10^{26}$ seconds), its effects can be considered separately from ${\cal L}_{adm}$.  

As stated earlier, we assume that the dark matter has antilepton number.  Dark matter that couples to the $Z$ boson is ruled out by direct detection, so we assume that it is a singlet under the Standard Model gauge groups.  Then, we have for benchmark $ii)$ of Eqn.~\ref{eqn:decays1} (in two component notation)
\bea
{\cal  L}_{decay} = y (H^\dag L) \chi + y (L^\dag H) \overline{\chi}\, \text{,}
\eea 
which couples the dark matter to the Higgs and lepton doublet, just like a right-handed neutrino.  Given the high mass of the dark matter, the equivalence theorem predicts that the branching ratios of $\chi$ are as in Eq.~\ref{eqn:decays1}, where the $L$ in the coupling determines if the antilepton produced is a muon or tau.  If instead, the Higgs doublet is a leptophilic Higgs, it decays into $H^0 \bar{\nu}, A^0 \bar{\nu}, H^- \ell^+$ in the branching ratio 1:1:2 with the motivated decays $(H^0, A^0) \to \tau \bar{\tau}$ and $H^- \to \tau \bar{\nu}.$  We do not consider the indirect detection signals of such a scenario in detail in this paper, but it presumably would give charge asymmetries somewhat intermediate between the benchmarks $i)$ and $ii)$ of Eqns.~\ref{eqn:decays0} and~\ref{eqn:decays1}.

If we consider dark matter transforming nontrivially under the Standard Model with nonzero hypercharge, then we notice an interesting tension between direct detection limits and charge asymmetry.  In particular, $Z$ boson exchange is ruled out by several orders of magnitude due to limits from direct detection experiments, for example from CDMS \cite{Ahmed:2009zw}, XENON100 \cite{Aprile:2011hi}, and EDELWEISS \cite{:2011cy}.  A commonly used dynamical technique to avoid such limits is to introduce a splitting so that the $Z$ exchange is inelastic (for e.g.~see \cite{Han:1997wn, Hall:1997ah}).  However, such inelastic splittings, which need to be $\gtrsim 100$ keV to avoid the limits, are usually generated by terms violating the dark matter number $U(1)$ symmetry.  For a violating mass term of this size,  the dark matter will have oscillated back into an equal distribution of dark and anti-dark matter before dark matter annihilations decouple, leaving only a symmetric component at late times \cite{Cohen:2009fz, Cai:2009ia}.  Thus, these terms would remove any charge asymmetric signals from present decays.  Such a tension makes the straightforward approach of realizing benchmark $i)$ and $iii)$ via the coupling
\bea
{\cal L}_{no\; good} = y (H^\dag \chi^c) e^c + y (\chi^{c\,\dag} H) \overline{e^c}
\eea
(where the dark matter transforms like the Higgs under $SU(2)_L \times U(1)_Y$) unviable.  

Adding one additional particle fixes this issue.  In particular, the dark matter can be taken to be a triplet with zero hypercharge, which has no $Z$ boson exchange at tree level and is safe from direct detection limits \cite{Cirelli:2005uq, Essig:2007az, Hisano:2011cs}.  Adding a Higgs triplet $\phi$ with hypercharge of one, we can write (for benchmarks $i)$ and $iii)$)
\bea
{\cal L}_{decay} &=& y (\phi^\dag \chi^c) e^c + y (\chi^{c\,\dag} \phi) \overline{e^c} \nonumber \\[-.2cm] 
\\[-.2cm]
&+& \mu H^T \phi H  + \mu H^\dag \phi^* H^*\nonumber \,\text{,}
\eea
where the last two terms (after electroweak symmetry breaking) mix the $\phi$ particle with minus one charge with the charged component of the Higgs, giving it the same decay channel.  Thus, depending on whether this Higgs is leptophilic or not, it would generate the decays of benchmark $i)$ or $iii)$.


\bibliography{admdecay}
\bibliographystyle{apsrev}

\end{document}